\documentclass[fleqn,12pt,twoside]{article}
\usepackage{units}
\usepackage{booktabs}
\usepackage{amsmath}
\usepackage{subfigure}
\usepackage{psfrag}
\usepackage{epsfig}
\usepackage{pstricks}
\usepackage{relsize}
\usepackage{amsmath}
\usepackage{amssymb}
\usepackage{wasysym}
\usepackage[headings]{espcrc1}
\readRCS
$Id: espcrc2.tex,v 1.2 2004/02/24 11:22:11 spepping Exp $
\usepackage{graphicx}
\usepackage[figuresright]{rotating}

\newcommand{\vect}[1]{\boldsymbol{#1}}

\renewcommand{\AmS}{{\protect\the\textfont2
  A\kern-.1667em\lower.5ex\hbox{M}\kern-.125emS}}

\title{Transport Simulation and Diffractive Event Reconstruction 
at the LHC\thanks{This work was supported in part by the ECO-NET Programme: 
                ``To Study the Forward Physics at the LHC and the Search 
                for the Higgs Boson in Diffraction''
                and Polish grants PBS-CERN/85/2006 and 319/N-CERN/2008/09/0}}

\author{R. Staszewski \address[IFJ]{ Institute of Nuclear Physics PAN,
ul. Radzikowskiego 152,\\
31-342 Krak\'ow, Poland.}  %
and
        J. Chwastowski \address[PK]{ Chair of Teleinformatics, \\
Faculty of Physics, Mathematics and Applied Computer Science,\\
                                    Cracow University of Technology,\\
                ul. Warszawska 24, 31-115 Krak\'ow, Poland.} \addressmark[IFJ]
}

\begin{document}

\maketitle

\begin{abstract}
The measurement of diffractively scattered protons in the ATLAS
Forward Physics detector system placed 220 m away from the ATLAS
interaction point is studied. A parameterisation of the scattered
proton transport through the LHC magnet lattice is presented. The
proton energy unfolding and its impact on the centrally produced
scalar particle mass resolution are discussed.
\end{abstract}

\section{Introduction}
\label{sec:intro}

Diffractive dissociation is one of the processes that can be studied
at the LHC. Diffractive physics is strong interaction physics
involving no exchange of quantum numbers other than those of the
vacuum. In experiment this leads to an obvious triggering scheme
relying on the rapidity gap method. However, one has to keep in mind
that the gap has a certain survival probability that depends on the
interaction type and the centre of mass energy. In addition, also the
diffractively scattered protons can be tagged.  Usually, the protons
scatter at small angles and in the collider environment they stay in
the beam pipe and travel through the magnet lattice of the machine. A
possible way to measure parameters of the diffractively scattered
proton trajectory is to use detectors placed inside the beam pipe, for
example by means of a roman pot station or a movable beam pipe
technique.  The ATLAS Collaboration plans to have proton tagging
stations placed symmetrically with respect to the Interaction Point
(IP) at the distances of 220 m and 420 m (AFP220 and AFP420) and 240 m
(ALFA). The ALFA (Absolute Luminosity For \textsc{Atlas}) \cite{ALFA}
stations at 240 m will be devoted to the absolute luminosity
measurement of the LHC at the ATLAS IP.  This measurement will rely on
the detection of the elastically scattered protons. The AFP
(\textsc{Atlas} Forward Physics) \cite{AFPLOI} stations will be used
for diffractive and $\gamma\gamma$ physics.

\section{Experimental Environment}
\label{sec:enviro}

Below, only the aspects concerning the AFP220 detectors are
discussed. In the measurement, the machine magnets play the role of
magnetic spectrometer.  Therefore, two detector stations are to be
placed around 220 m, namely one at 216 and one at 224 m.  Each station
will be equipped with position sensitive and triggering detectors,
horizontally inserted into the LHC beam pipe. The position sensitive
detectors will consist of 10 layers of the silicon 3D detectors
\cite{silicon} to measure the scattered proton trajectory. The
position measurement resolutions are assumed to be $\sigma_x = 10\
\mum$ and $\sigma_y = 40\ \mum$ in the horizontal and vertical
direction, respectively. This will allow the measurement of the
particle trajectory position and direction.  Additionally, the
stations will contain very fast timing detectors with picosecond
resolution.  They will measure the scattered proton time of flight and
indicate the interaction vertex longitudinal coordinate, $z$, with
resolution of the order of several millimeters.
In the following calculations, a reference frame with the $x$--axis pointing towards
the accelerator centre, the $y$--axis pointing upwards and the
$z$--axis along one of the beams was used.

The scattered proton can be described at the interaction point in
several equivalent ways, each useful in a different case: $(p_x, p_y,
p_z)$, $(E, x_0',y_0')$, or $(E, \vect{p_T})$, $$ E = \sqrt{m^2 +
\vect{p}^2} \qquad x_0' = \frac{p_x}{p_z} \qquad y_0' =
\frac{p_y}{p_z} \qquad $$ where $E$ and $m$ are the proton energy and
mass, $\vect{p} = (p_x, p_y, p_z)$ is the proton momentum and
$\vect{p_T} = (p_x,p_y)$ denotes the proton transverse
momentum. Useful variables are the proton energy loss $\Delta E =
E_0-E$ ($E_0$ is the incident beam energy), the reduced proton energy
loss, $\xi = \Delta E/E_{0}$, and the four-momentum transfer between
the incident and scattered proton, $t$.

There are several programs on the market calculating proton trajectories
 through the magnets. In the following the FPTrack program
\cite{FPTrack} was used. This program computes the positions of particles
using the LHC optics files which describe the magnetic fields,
positions and apertures of the LHC lattice. These files were produced
with help of a principal beam transport program Mad-X \cite{MadX} by
the LHC optics group \cite{LHCoptics}.
\begin{table}[h]
  \centering
  \caption{The LHC beam and the crossing region parameters at the ATLAS IP.}
  \label{tab:beam_parameters}
  \[
  \begin{array}{c c c c}
    \toprule
    \textbf{Parameter} & \textbf{Beam} & \textbf{Crossing region} \\ 
    \midrule
    \sigma_{x_0}  & 16.6\ \mum   & 11.7\ \mum    \\
    \sigma_{y_0}  & 16.6\ \mum   & 11.7\ \mum    \\
    \sigma_{z_0}  & 75\ \mm      & 53\ \mm       \\
    \sigma_{x_0'} & 30.2\ \murad & -  \\
    \sigma_{y_0'} & 30.2\ \murad & -  \\
    \sigma_{E_0}  & 0.77\ \GeV   & -  \\
    \bottomrule
  \end{array}
  \]  
\end{table}
%

The interaction vertex is
described by its coordinates ($x_0$, $y_0$, $z_0$). These coordinates
have Gaussian distributions with zero average values and dispersions:
$\sigma_{x_0}$, $\sigma_{y_0}$, $\sigma_{z_0}$, respectively. In the
simulation, the beam particle energy and its momentum direction were
generated according to Gaussian distributions with appropriate means
and the dispersions: $\sigma_{E_0}$ for the energy and $\sigma_{x_0'},
\sigma_{y_0'}$ for the momentum in $(x,z)$ and $(y,z)$ planes,
correspondingly.
Values of these parameters for the nominal 7 TeV beam energy and standard LHC 
optics are listed in Table \ref{tab:beam_parameters}. 
The calculations were performed for both beams: $beam1$ that
performs the clockwise motion and $beam2$ which does the counter
clockwise rotation.
\begin{figure}[ht]
  \centering 
\vspace{-10mm}
  \psfrag{XAXIS}[l]{\hspace{-6mm}\raisebox{-9mm}{\large $x$ [mm]}}
  \psfrag{YAXIS}[l]{\hspace{-6mm}\raisebox{10mm}{\large $y$ [mm]}}
  \subfigure{
    \includegraphics[width=75mm,height=75mm]{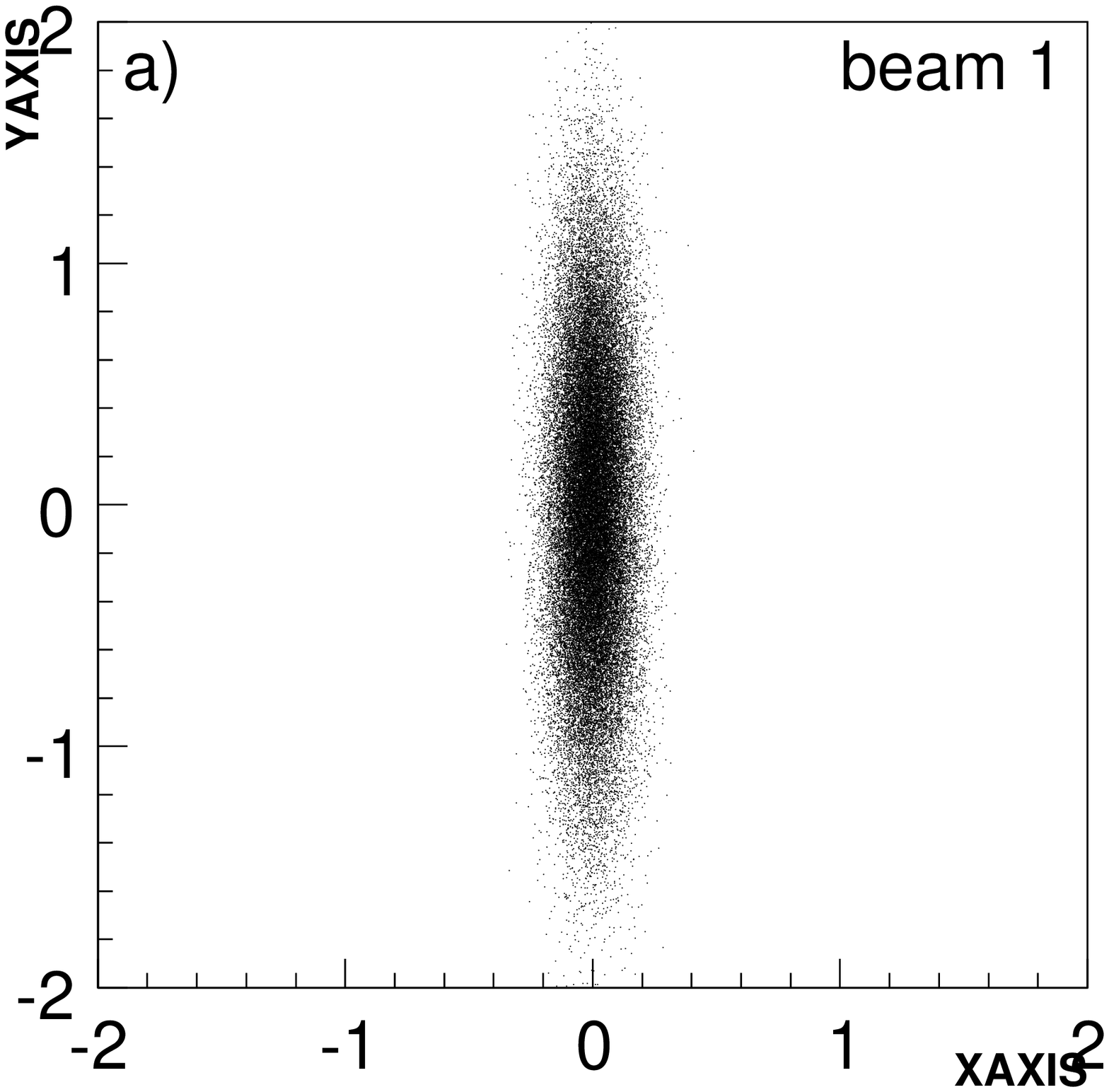}
  }
  \subfigure{
    \includegraphics[width=75mm,height=75mm]{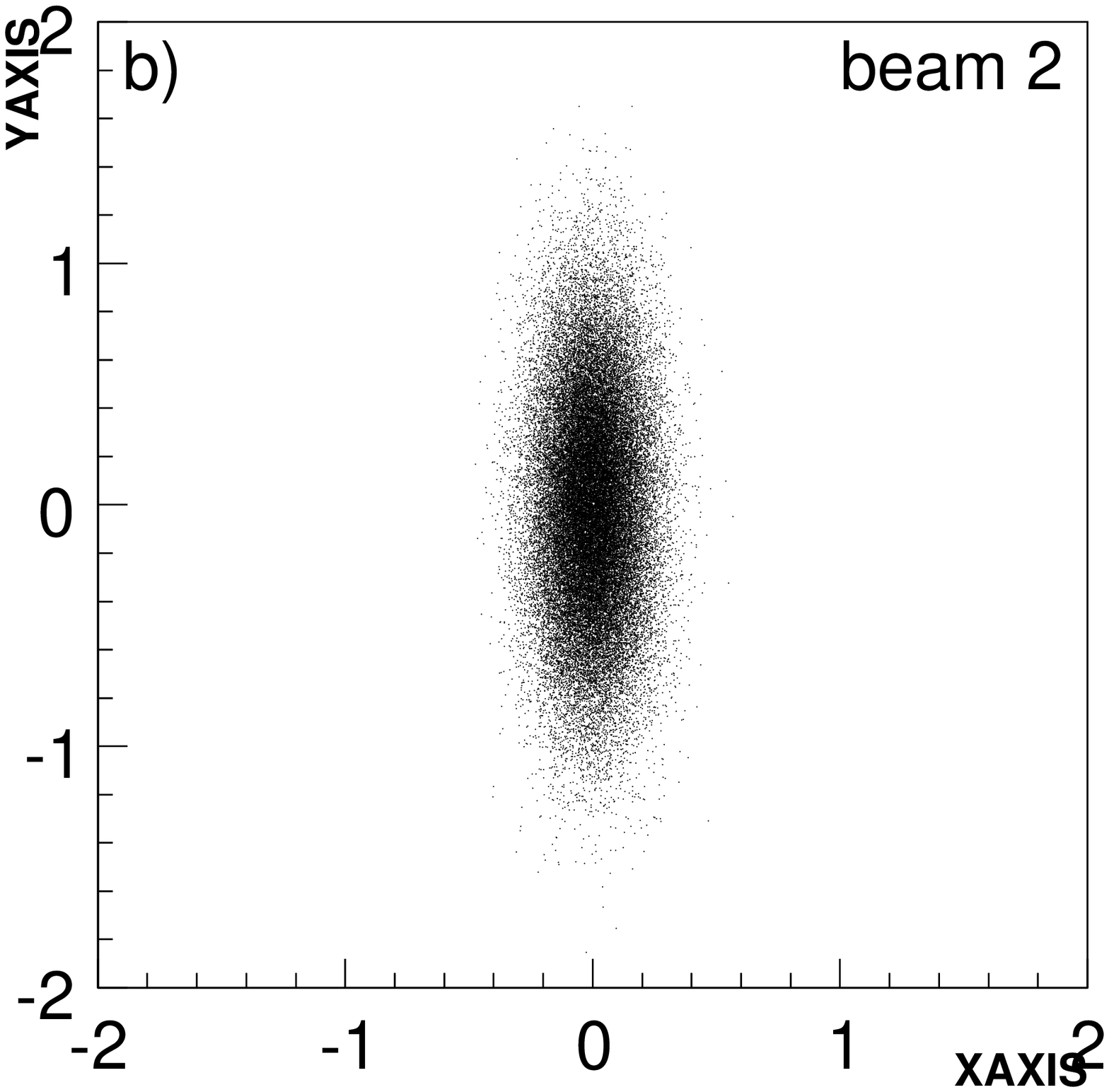}
  }
\vspace{-7mm}
  \caption{The LHC beam profiles at 216 meters from ATLAS IP for $beam1$
    (a) and $beam2$ (b), for the 7 TeV LHC optics.}
  \label{fig:beam_profiles}
\end{figure}

Figure \ref{fig:beam_profiles} shows the LHC beam profiles in the
($x$,$y$) plane at 216 m away from the ATLAS IP obtained with the
FTPtrack program for both beams and the 7 TeV LHC optics. As can be
observed, the beams are much wider in the vertical direction than in
the horizontal one.

Table \ref{tab:beam_fit_params} summarizes the beams spreads following
from a two dimensional Gaussian fit to the beam profiles shown in
Fig. \ref{fig:beam_profiles}. The (10 -- 15)$\sigma$ beam envelope
gives a natural limit for the distance between the detector frame and
the beam centre.  In the horizontal direction this corresponds to
about 1 -- 2 millimeters.  Obviously, this distance plays a crucial
role for the diffractively scattered proton detection and hence, for the 
experimental apparatus acceptance.
\begin{table}[h]
  \centering
  \caption{The LHC beam spreads from a Gaussian fit to the beam profiles at 
    216~m.} 
  \label{tab:beam_fit_params}
  \[
  \begin{array}{c c c c}
    \toprule
    \textbf{Parameter} & \textbf{Beam1} & \textbf{Beam2}\\ 
    \midrule
    \sigma_{x_{216}}  & 88\ \mum   & 121\ \mum    \\
    \sigma_{y_{216}}  & 569\ \mum   & 421\ \mum    \\
    \bottomrule
  \end{array}
  \]  
\end{table}

It is important to see in which range of the energy, $E$, and the
transverse momentum, $p_T = |\vect{p_T}|$, the detector can measure
protons. The geometric acceptance for fixed $E$ and $p_T$ values was
defined as the ratio of the number of protons that crossed the
detector to the total number of scattered protons with a given $E$ and
$p_T$. Only the effects of the beam pipe aperture and the distance
between the detector and the beam centre were taken into account.

\begin{figure}[htbp]
  \centering
  \psfrag{XAXIS}[l]{    \hspace{20mm}{\large $p_T$ [GeV/c]}}
  \psfrag{YAXIS}[l]{    \hspace{6mm}\raisebox{5mm}%
    {\large $\Delta E$ [GeV/c$^2$]}}
  \psfrag{ZAXIS}[l]{    \hspace{2mm}{\large Acceptance [\%]}}
\vspace{-1.5cm}
    \includegraphics[width=10cm]    {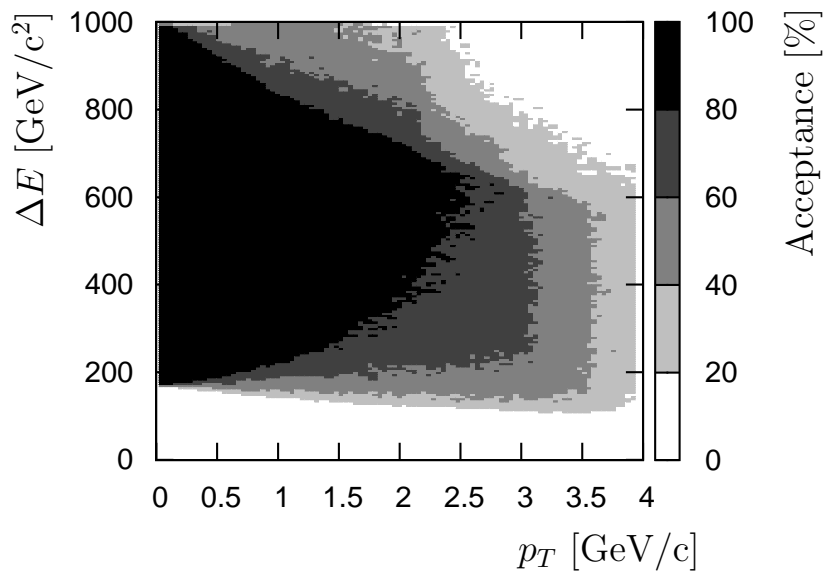}
\vspace{-15mm}
    \caption{The geometrical acceptance of the detector placed in the
	LHC $beam1$ at 216 m away from the IP as a function of the
	proton energy loss $(\Delta E)$ and its transverse momentum
	$p_T$ for a 3 mm distance between the beam centre and the detector 
	active edge.}
  \label{fig:acceptance}
\end{figure}

Figure \ref{fig:acceptance} depicts the geometric acceptance as a
function of $E$ and $p_T$ for $beam1$ for the standard 7 TeV LHC
optics. The acceptance is above 80\% in the region limited by:
$$ 200<\Delta E<1000\ [\GeV],\ 0<p_T< 2.5\ [\GeV/c] $$ 
which corresponds to
$$ 0.03<\xi< 0.14 ,\  -6.5 < t < 0\ [\GeV^2/c^2]. $$ 

When the geometrical acceptance requirement is lowered to 60\%, this
results in a wider range of the accepted proton energies and
transverse momenta.  The range enlargement is seen
(c.f. Fig. \ref{fig:acceptance}) for $ 200 < \Delta E < 600$ GeV and 
$p_T<3$ GeV/c. This gives the limits: $$ 0.03 < \xi < 0.14,\ -10 < t <
0\ [\GeV^2/c^2].$$

The presence of two detector stations on each side of the ATLAS
detector allows the measurement of the proton trajectory elevation
angles, $x\,'$ and $y\,'$, in the ($x$,$z$) and ($y$,$z$) planes,
respectively. From the FPTrack calculations, it follows that the
position and slope of the trajectory at the detector in one transverse
direction is independent of those in the other direction, i.e. $x$ and
$x'$ values do not depend on $y_0$ and $y_0'$ and vice versa. This is
a reflection of a negligible role of the sextupole and higher order
magnetic fields in the standard LHC optics between the ATLAS IP and
the AFP220 stations.
\begin{figure}[htbp]
  \centering
\vspace{-10mm}
  \psfrag{XAXIS}[l]{\hspace{15mm} \large $x$ [mm]}
  \psfrag{YAXIS}[l]{\hspace{15mm} \large $x'$[$\mu$rad]}
  \subfigure
    {
    \includegraphics[width=75mm, height=75mm]    {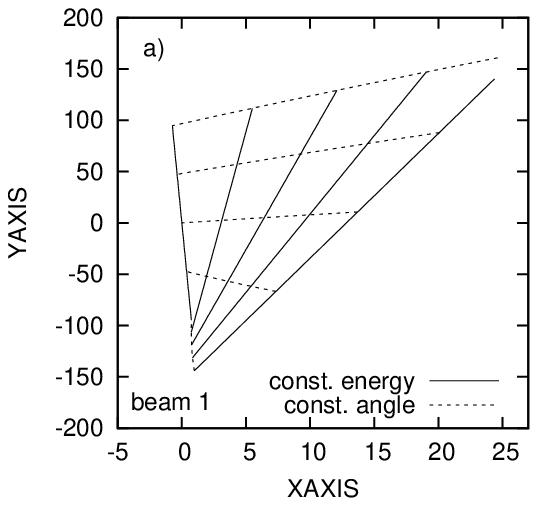}
  }
  \subfigure
    {
    \includegraphics[width=75mm, height=75mm]    {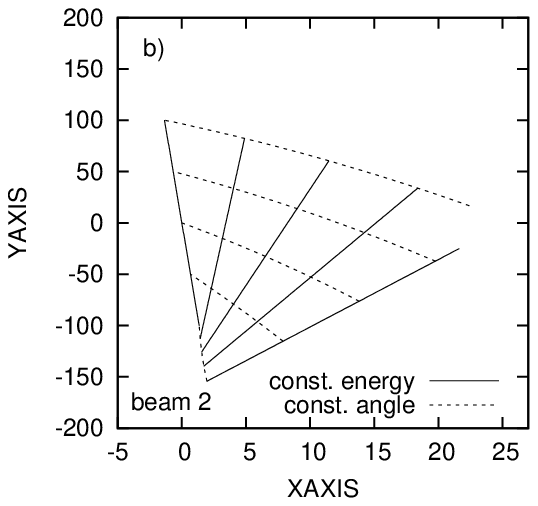}
  }
\vspace{-15mm}
  \caption{The $x$-direction chromaticity plots for the LHC $beam1$
  (a) and $beam2$ (b) for the standard 7 TeV LHC optics.  The lines of
  constant energy correspond to 7000, 6825, 6650, 6475, 6300 GeV from
  left to right, respectively. The angles were changes from
  -400~$\mu$rad to 400 $\mu$rad (from top to bottom).}
  \label{fig:chromaticity_x}
\end{figure}

To illustrate how the proton trajectory positions and slopes measured
by the detectors depend on the proton energy and its trajectory slopes
at the IP the chromaticity plots were prepared.  The plots shown in
Figure \ref{fig:chromaticity_x} were devised by plotting in the
($x$,$x'$) plane the lines corresponding to the constant $E$ and
$x_0'$ at the IP.
 
The chromaticity plots indicate few things. Firstly, there is a
non-negligible difference between both beams. Therefore, properties of
both have to be studied. Secondly, the grids created by the energy and
angle iso-lines do not fold. Hence, it is possible to obtain the
energy and the transverse momentum of a proton from the measurements
of the proton trajectory in both stations. In particular, assuming a
fixed interaction vertex position (no smearing) the energy of a proton
at the IP can be deducted solely from the measured $x$ and $x'$
values.

\section{Transport Parameterisation}
\label{sec:tranport}

In order to unfold the proton energy from the detector measurements a
parameterisation of the FPTrack transport calculations was prepared.
The aim was to describe the FPTrack results analytically. It was
requested that:
\begin{itemize}
\item the parameterisation has a simple functional form,
\item the parameterisation precision has an accuracy which is better
      than the assumed detector spatial resolutions.
\end{itemize}

To find the parameterisation form, events uniformly distributed over
the $(E, x_0',y_0',$ $x_0, y_0, z_0)$ space were
generated. Subsequently, these events were used in the FPTrack
transport calculations. The transport results were the input data to
the parameterisation search procedure. It was found that the
following parameterisation fulfill the requirements outlined above
well:
\begin{equation}
  \alpha  = A_\alpha + \alpha_0'B_\alpha + \alpha_0C_\alpha + 
  \alpha_0'z_0D_\alpha + z_0F_\alpha,
  \label{eq:param1}
\end{equation}
\begin{equation}
  \alpha\,' = A_{s\alpha} + \alpha_0'B_{s\alpha} + \alpha_0C_{s\alpha}
  + \alpha_0'z_0D_{s\alpha} + z_0F_{s\alpha},
  \label{eq:param2}
\end{equation}
where $\alpha = \{x,y\}$, $s\alpha$ denotes the slope either in $x$ or
$y$ direction and all the capitalised symbols are polynomials of
energy, i.e.:
\begin{equation}
  A_\alpha = a_\alpha^{(0)} + a_\alpha^{(1)}E + a_\alpha^{(2)}E^2 + 
  a_\alpha^{(3)}E^3 + a_\alpha^{(4)}E^4,
  \label{eq:parameterisation_a}
\end{equation}
\begin{equation}
  C_{s\alpha} = c_{s\alpha}^{(0)} + c_{s\alpha}^{(1)}E + c_{s\alpha}^{(2)}E^2 +
  c_{s\alpha}^{(3)}E^3.
  \label{eq:parameterisation_b}
\end{equation}
\noindent
The values of all the coefficients were found by fitting the 
formulae to the FTPtrack calculations for simulated events. 

\begin{figure}[ht]
  \centering
\vspace{-10mm}
  \psfrag{XAXIS}[l]{\hspace{-10mm}\raisebox{-10mm}{\large $\Delta y$ [$\mu$m]}}
  \psfrag{YAXIS}[l]{\hspace{-7mm}\raisebox{15mm}{\large Events}}
  \subfigure{
    \includegraphics[width=74mm]{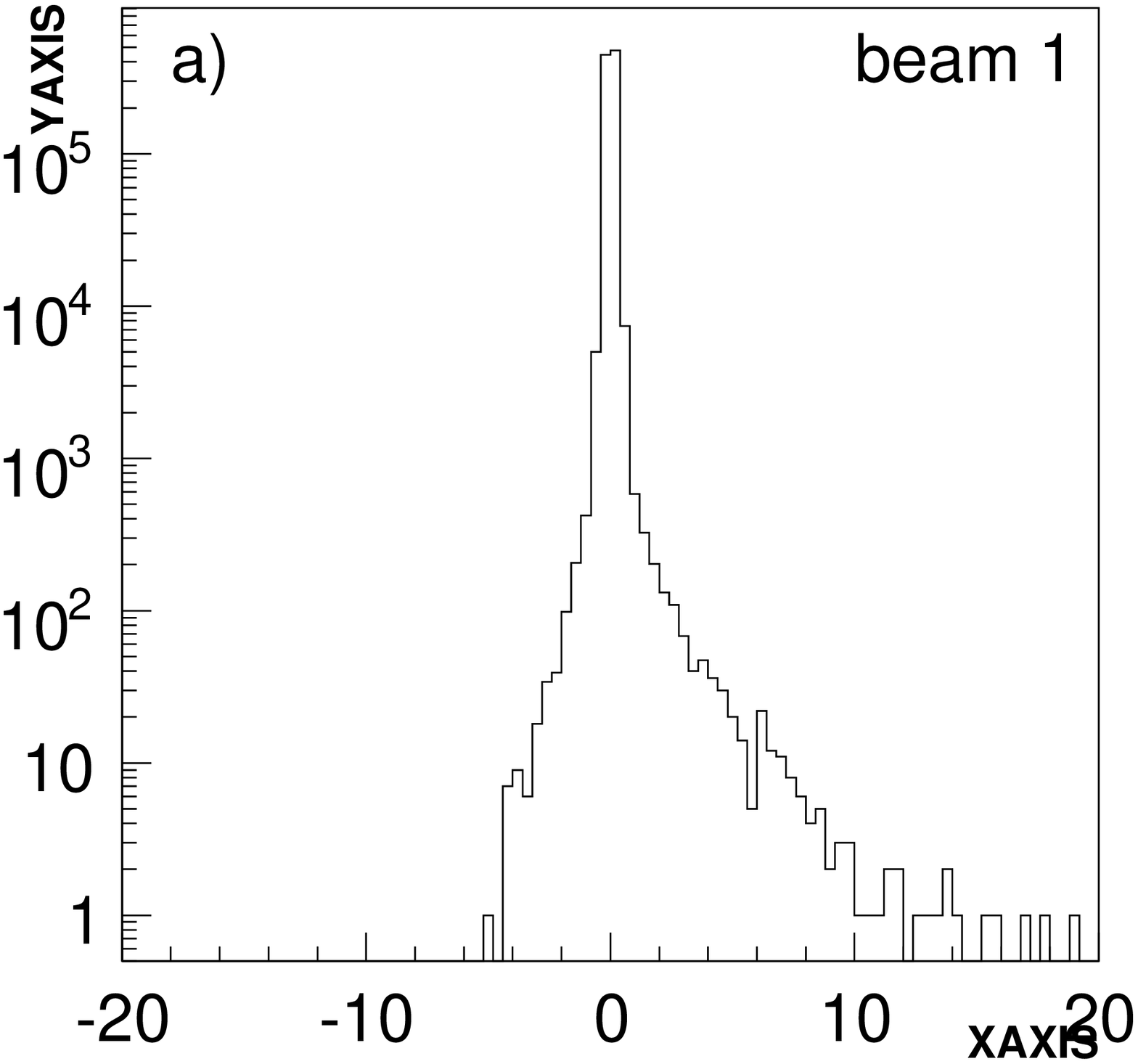}
  }
 \psfrag{XAXIS}[l]{\hspace{-15mm}\raisebox{-10mm}{\large $\Delta y\,'$ [nrad]}}
  \psfrag{YAXIS}[l]{\hspace{-7mm}\raisebox{15mm}{\large Events}}
  \subfigure{
    \includegraphics[width=74mm]{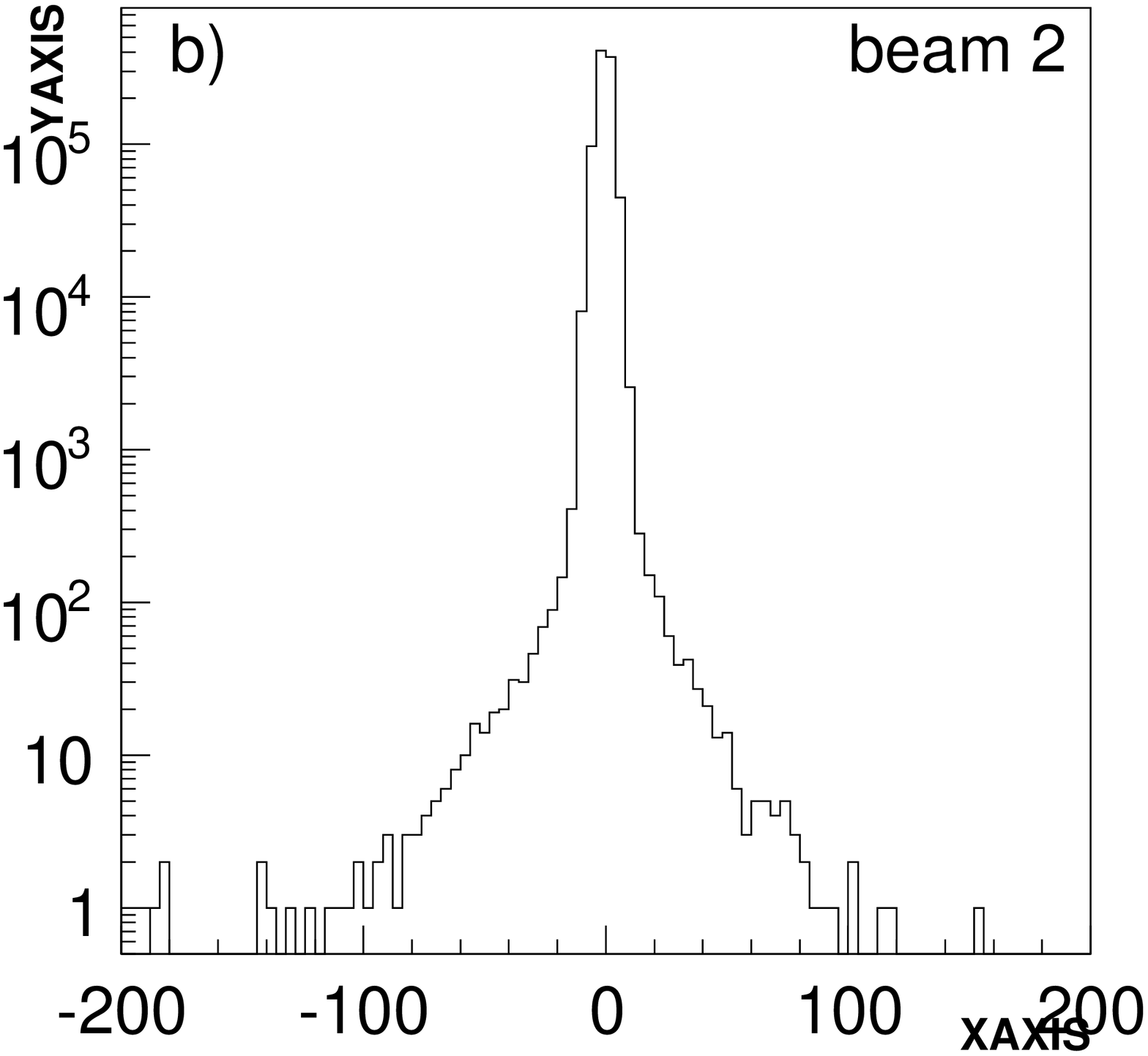}
  }
\vspace{-7mm}
  \caption{The parameterisation accuracy estimation examples (see text). 
    Pictures show the uncertainty on the $y$ (a) and $y'$ (b) 
    parameterisations of the $beam1$ transport. }
  \label{fig:parameterisation_error}
\end{figure}

The accuracy of the method was estimated by plotting the difference
between the value given by the parameterisation and that given by the
FPTrack calculation.  The accuracy of the position parameterisation
was found to be of the order of a micrometer which is 10 times less
than the assumed detector resolution in the horizontal plane. The
difference between FPTrack and the parameterisations for the
trajectory angles was found to be limited to about 50 nanoradians. One
has to remember that the average multiple Coulomb scattering angle was
estimated to be about 500 nrad. An example of the parameterisation
accuracy is presented in Fig. \ref{fig:parameterisation_error}.  In
this figure the distributions of $\Delta y = y_{param}-y_{FPTrack}$
and $\Delta y' = y_{param}'-y_{FPTrack}'$ are shown for single
diffractive events generated with PYTHIA~\cite{pythia}.  The accuracy
estimations for these quantities are displayed since they represent
the worst precision cases.  Nevertheless, the results are well
confined within the ranges given by the detector resolutions. This
confirms the parameterisation quality.

One should note that the procedure outlined above can be easily repeated. 
In particular, it can be applied to the files describing the actual
LHC collision optics used for experimental runs.
\section{Event Reconstruction}
\label{sec:reco}

Since there is a correlation between the proton momentum and the
measured position of the proton at the AFP220, the reconstruction of
the proton properties at the interaction vertex from the measured
coordinates of the proton trajectory at the AFP220 is possible.

A proton at the interaction vertex is described by six independent
variables: $E$, $x_0'$, $y_0'$, $x_0$, $y_0$ and $z_0$. The detectors
deliver two pairs of transverse coordinates separated in longitudinal
direction by a fixed distance of about 8 meters. In general, the
unfolding problem is an ill-stated one. In the present case it
requests the inversion of the 6 to 4 mapping and its solution is only
possible with help of additional assumptions. The simplest one is the
assumption of a fixed position of the interaction vertex.  In the
following the positions $x_0 = y_0 = 0$ and $z_0 = 0$ or $z_0 =
216-c\tau$ were chosen, where $\tau$ is the proton time of flight.

To make the measurement simulation as close to reality as possible the
detector effects were taken into account. Protons traversing the
detector station undergo multiple Coulomb scattering in the frame
and the detector materials. The simulation of the proton trajectory
position measurement also takes into account the assumed detector
resolutions. These effects may lead to a considerable change of the
proton trajectory parameters (slopes and positions). Hadron
interactions in the detector or its frame were neglected as they are
not important for the present study.

A simple and fast method of the proton energy unfolding from the
detector measurement is proposed. This method uses the assumption that
the values actually measured are equal to those delivered by the
parameterisation.  This allows to calculate $x_0'$ from
eqs.~(\ref{eq:param1})~and~(\ref{eq:param2}). Since both equations are
considered for the same particle, they should give equal
values. Hence, after simple algebra one gets:
\begin{multline}
  (x  -  A_x  -  F_x z_0  -  x_0 C_x)\cdot
  (B_{sx} + z_0 D_{sx}) =\\=
  (x' - A_{sx} - F_{sx} z_0 - x_0 C_{sx})\cdot
  (B_x  +  z_0 D_x)
\end{multline}
where all capitalised symbols are described by eqs. \ref{eq:parameterisation_a}
and \ref{eq:parameterisation_b}.

The solution of the above equation is equivalent to finding the zero of
the function $f(E)$ given below:
\begin{multline}
f(E) =  
  (x  -  A_x  -  F_x z_0  -  x_0 C_x)\cdot
  (B_{sx} + z_0 D_{sx}) -\\
  (x' - A_{sx} - F_{sx} z_0 - x_0 C_{sx})\cdot
  (B_x  +  z_0 D_x).
\label{eq:fe}
\end{multline}
It was observed that for obtained parameterisation and simulated
events the function $f(E)$ has only one zero. Therefore, the equation
$$ f(E) = 0 $$
can be easily solved numerically using for example the bisection method
\cite{numerical_recipes}.

The energy unfolding procedure was tested using the same PYTHIA 
generated data sample. The proton energy was reconstructed with the help
of the different additional assumptions listed below:
\begin{itemize}
\item the ``measured'' trajectory coordinates were smeared according to the
 detector resolution, 
\item the interaction vertex transverse position was exactly known,
\item the interaction vertex longitudinal position was exactly known.
\end{itemize}
 
\begin{figure}[h]
  \centering \psfrag{XAXIS}[l]{\hspace{-37mm} \raisebox{-12mm}
{\large  Proton energy [GeV]}}
 \psfrag{YAXIS}{\hspace{-45mm} \raisebox{5mm}{\large Energy rec. resol. [GeV]}}
  \includegraphics[width=12cm] {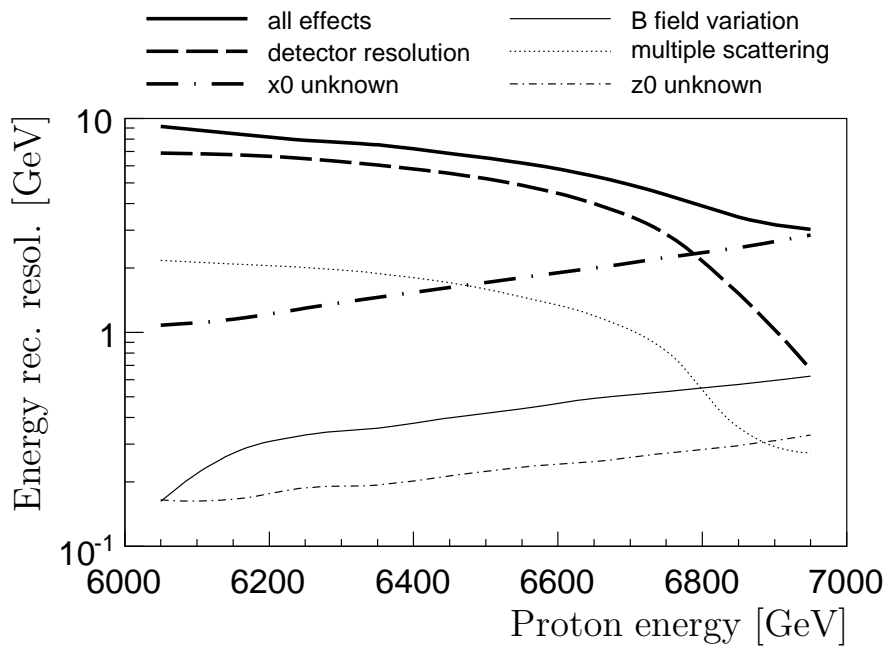} 
\caption{The proton
  energy reconstruction resolution for $beam1$ as a function of its
  energy. The overall resolution is marked with the thick solid line,
  the influence of: the detector spatial resolution -- the thick dashed
  line, the vertex position in the transverse plane -- the thick
  dash-dotted line, the multiple Coulomb scattering -- the dotted
  line, the vertex position along the beam axis -- the dash-dotted line, 
  the magnetic field variation -- the solid line.}
\label{fig:E_resolution}
\end{figure}

The results are presented in Figure \ref{fig:E_resolution}.  The
energy reconstruction resolution (the thick solid line) decreases from 9 GeV for
6000 GeV protons to about 3 GeV for 7000 GeV protons. It is dominated
by the detector spatial resolution which influence, marked with the thick
dashed line, decreases with proton energy from about 7 GeV to about 1
GeV within considered energy range. Also, the impact of the multiple
Coulomb scattering (dotted line) gets smaller with increasing proton
energy. Its contribution to the resolution is about 2.5 GeV at the
maximum.  For proton energies greater than 6800 GeV the uncertainty on
the interaction vertex position in the transverse plane (the thick
dash-dotted line) dominates the energy reconstruction resolution. The
influence of the interaction vertex position along the longitudinal
axis (the dash-dotted line) is small in the whole energy range
discussed.

Since the scattered protons can traverse the whole beam pipe volume
the influence of the possible imperfections of the magnetic fields was
studied. The magnetic fields of the lattice were varied by $\pm
1\permil$ of their nominal values. It should be pointed out that
assumed variation is about a factor of 10 larger that the machine
accepted and about 50 times larger than the measured values
\cite{field_quality} of the higher multipoles at the reference radius
of 17~mm away from the beam pipe centre. Variation of the magnetic
field values gives a small contribution to the energy reconstruction
resolution and for 6000 GeV protons it is about 0.3 GeV and increases
to 0.7 GeV at 7000 GeV. This contribution is marked with the 
solid line in Fig. \ref{fig:E_resolution}. The other effect of the
variation of magnetic fields is the offset of the scattered proton
reconstructed energy.  This offset, on the absolute value, decreases
linearly from about 1.3 GeV to approximately 0.1 GeV for proton
energies between 6000 and 7000 GeV.

Another important experimental factor is the detector alignment.  It
is required that the detector stations will be able to measure the
scattered proton trajectory elevation angles with precision of about 1
$\mu$rad. This implies a 10 $\mu$m precise alignment. It turned out that
the 10 $\mu$m misalignment of the stations results an offset of the
reconstructed proton energy. This offset has the largest value of
about 5 GeV for protons of 6000 GeV energy and decreases to zero with
proton energy increasing to 7000 GeV.

\section{Central Exclusive Production (CEP)}
\label{sec:cep}

The AFP detectors can be used to measure the exclusive central
production of scalar $J^{PC}=0^{++}$ particles (for example the Higgs
boson or some supersymmetrical particles). The central production can
be viewed as a two stage process.  In a first step each of the
incident protons emits a color singlet object.  Subsequently, these
objects interact with each other giving a centrally produced
system. The incoming protons remain intact, traverse the magnetic
lattice of the machine inside the beam pipe and can be detected in the
AFP detectors.  The centrally produced system decays into the ATLAS
main detector.  Hence, this gives a unique possibility to measure all
the particles belonging to the final state (a completely exclusive
event measurement).
\begin{figure}[h]
  \centering
  \psfrag{XAXIS}{\hspace{-40mm}\raisebox{-5mm}
    {\large Missing mass [GeV/c$^2$]}}
  \psfrag{YAXIS}{\hspace{-25mm}\raisebox{8mm}%
    {\large Acceptance [\%]}}
    \includegraphics[width=9cm]  {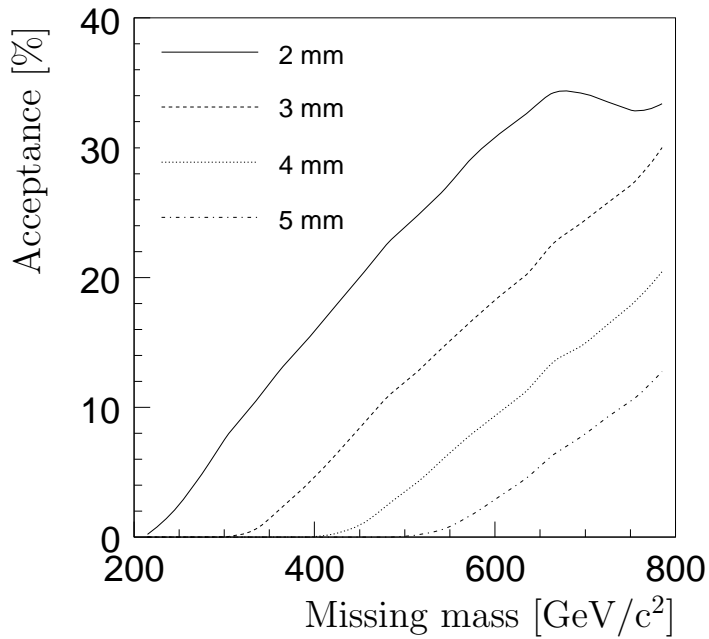}
    \caption{The geometrical acceptance of detectors at 216 m away
      from the ATLAS Interaction Point for Central Exclusive Production process
      as a~function of the produced particle mass. The solid lines depicts the
      acceptance for active detector region at the 2 mm distance from the
      beam centre. The dashed, dotted and dash-dotted lines mark the acceptance
      curves for the 3, 4 and 5 mm distance, respectively.}
    \label{fig:Mx_acceptance} 
\end{figure}

Such events were simulated in a simplified way. In the generation the
four momentum transfer, $t$, and the reduced proton energy loss,
$\xi$, were distributed according to $e^{-bt}$ with $b = 6\,\GeV^{-2}$
and $\xi^{-1}$, respectively. Later, the proton transport to the
AFP220 detectors was simulated using the FPTrack calculations.

In Figure \ref{fig:Mx_acceptance} the geometrical acceptance for
different masses of the centrally produced system for various
distances between the detector edge and the beam centre is shown.  As
expected, the geometrical acceptance strongly depends on this distance
and for a realistic distance of 3 mm (marked with the dashed line in
Figure \ref{fig:Mx_acceptance}) it varies between 0 and 30\% for masses
changing from 300 to 800 GeV.

Next, the mass of the centrally produced system was estimated using
the detector measurements and the proton energy reconstruction
described in section \ref{sec:reco}.  For the Central Exclusive
Production process the produced system mass determination from the
reduced energy losses of both protons, $\xi_1$ and $\xi_2$, is
possible via \cite{Albrow:2000na}: 
$$M_x = \sqrt{s\cdot\xi_1\cdot\xi_2},$$ 
where $s$ is the centre of mass energy squared.

\begin{figure}[h]
  \centering \psfrag{XAXIS}{\hspace{-40mm}\raisebox{-5mm} 
{\large  Missing mass [GeV/c$^2$]}} 
\psfrag{YAXIS}{  \hspace{-50mm}\raisebox{5mm} 
{\large Mass rec. resolution  [GeV/c$^2$]}} 
\includegraphics[width=12cm] {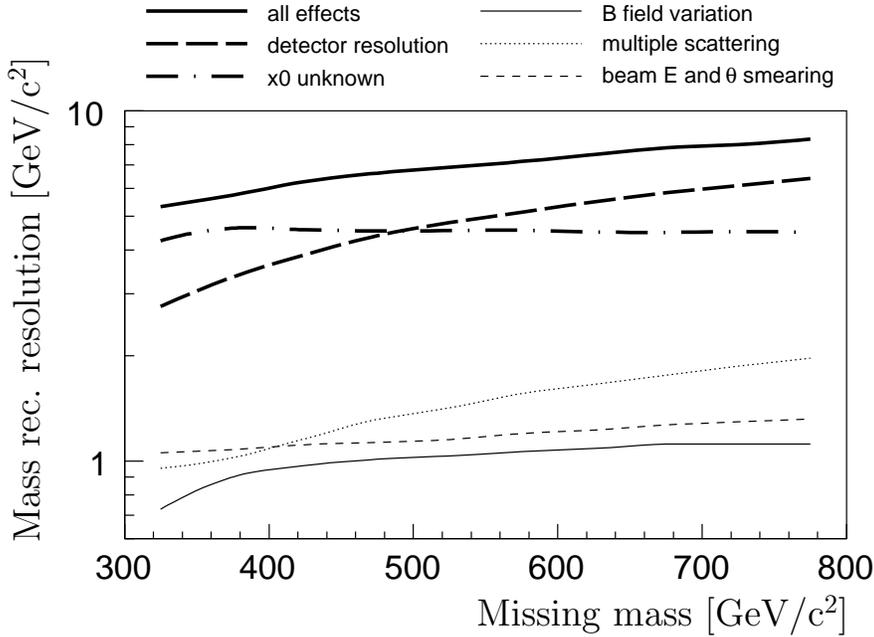}
  \caption{The centrally produced particle mass resolution determined
  with outgoing protons as a function of the particle mass. The
  overall mass reconstruction resolution is marked with the thick
  solid line, the influence of: the detector spatial resolution -- the
  thick dashed line, the vertex position in the transverse plane -- the
  thick dash-dotted line, the multiple Coulomb scattering -- the
  dotted line, the magnetic field variation -- the solid line, the beam
  energy and the proton direction angular spreads -- the dashed line.}
  \label{fig:Mx_resolution}
\end{figure}

The mass reconstruction resolution as a function of the centrally
produced system between 300 and 800 GeV is shown in Figure
\ref{fig:Mx_resolution} for the 3 mm distance between the detector and
the beam centre.  The impact of several experimental factors is also
depicted in this figure.  The mass reconstruction resolution, after an
initial jump at the acceptance edge, very slowly increases from 5 to 8
GeV with increasing value of the produced system mass. The influence
of the multiple scattering, the beam energy variation and the proton
direction angular spread is small and below 2 GeV.  In fact, the
resolution value is dominated by two factors.  First one is the
detector spatial resolution which gives the contribution ranging
between 2 and 6 GeV and which dominates for masses above 500 GeV. The
second one is the uncertainty on the $x_0$ coordinate of the
interaction vertex, whose influence practically does not depend on the
produced system mass and which is the most important factor for masses
below 500 GeV. The field imperfections, estimated as described
previously, have a small influence on the reconstructed mass
resolution which is about 0.7 GeV at 300 GeV and saturates at the
value of approximately 1 GeV at 500 GeV. Also, in this case the field
variation resulted the mass offset which is about 1 GeV in the
considered mass range. The impacts of the interaction vertex position
and that of the detector misalignment, not shown in Figure
\ref{fig:Mx_resolution}, are small and below 0.5$\permil$ of the
produced mass value.  The detector misalignment introduces the
reconstructed mass shift which almost linearly increases with the
produced mass value from about 2.5 GeV at 300 GeV to 6.5 GeV at 800
GeV.

\section{Summary and Conclusions}  

A parameterisation of the proton transport through the magnet lattice of the
LHC was devised. This parameterisation has a simple functional form and
enables fast and easy calculations.  

A proton energy unfolding procedure from the proton trajectory
position measurements was prepared. This procedure allows the
reconstruction of the scattered proton energy. The procedure was used
to reconstruct the missing mass of the centrally produced scalar
system.  The missing mass reconstruction resolution weakly depends on
the produced mass and reaches about 8 GeV at the mass value of 800
GeV.

The proton energy unfolding procedure can be used for the first level
triggering of the apparatus at the LHC environment.

\section*{Acknowledgments}
We are grateful to P. Bussey, A. Kupco, C. Royon, A. Siemko and  J. Turnau 
for many discussions, useful remarks and help.

\end{document}